%% file: main.tex
\documentclass[sigconf]{acmart}
\makeatletter
\def\@ACM@checkaffil{% Only warnings
    \if@ACM@instpresent\else
    \ClassWarningNoLine{\@classname}{No institution present for an affiliation}%
    \fi
    \if@ACM@citypresent\else
    \ClassWarningNoLine{\@classname}{No city present for an affiliation}%
    \fi
    \if@ACM@countrypresent\else
        \ClassWarningNoLine{\@classname}{No country present for an affiliation}%
    \fi
}
\makeatother
\usepackage{etoolbox} % For compile control
\AtBeginDocument{%
  \providecommand\BibTeX{{%
    \normalfont B\kern-0.5em{\scshape i\kern-0.25em b}\kern-0.8em\TeX}}}

\setcopyright{acmcopyright}
\copyrightyear{2022}
\acmYear{2022}
\makeatletter
\renewcommand\@formatdoi[1]{\ignorespaces}
\makeatother

\acmConference[KDD '22 DLG Workshop]{}{2022}{Washington, DC, USA.}
\acmBooktitle{KDD '22 Deep Learning on Graphs Workshop}
\acmPrice{15.00}
\acmISBN{}

\usepackage{hyperref}
\usepackage{url}

\usepackage{multicol}     % multi-columns

\usepackage{amsmath}
\usepackage{graphicx}
\usepackage{subcaption}
\usepackage{listings}
\usepackage{float}
\usepackage{enumitem}
\usepackage{xspace}
\usepackage{multirow}
\usepackage{array}
\usepackage{tabularx}
\usepackage{tabu}
\usepackage{bm}

\usepackage{booktabs, makecell}
\usepackage{xcolor,colortbl}
\usepackage{cleveref}
\usepackage[english]{babel}
\usepackage{pifont} 
\usepackage{soul}
\usepackage{slashbox}
\usepackage{xfrac}
\usepackage{wrapfig}
\usepackage{mdwlist}
\usepackage{arydshln}

\usepackage{adjustbox}

\usepackage{cellspace}
\usepackage{siunitx}
\usepackage{comment}

\usepackage{tikz}
\tikzstyle{xaxis-fill}=[rectangle,fill=black,text=white,minimum width=53em, minimum height=.65em, label={center:\scriptsize \bf \textcolor{white}{\textsf{1-Hop Sensitive Attribute Homophily}}}]
\usepackage{titlesec}
\titlespacing*{\subsubsection}{0pt}{.3\baselineskip}{.3\baselineskip}
\titlespacing*{\subsection}{0pt}{.4\baselineskip}{.3\baselineskip}
\titlespacing*{\section}{0pt}{.6\baselineskip}{.3\baselineskip}
\usepackage[linesnumbered,ruled,vlined]{algorithm2e}

\usepackage{tikz}

\crefformat{section}{\S#2#1#3} % see manual of cleveref, section 8.2.1
\crefformat{subsection}{\S#2#1#3}
\crefformat{subsubsection}{\S#2#1#3}
\newcolumntype{L}[1]{>{\raggedright\let\newline\\\arraybackslash\hspace{0pt}}m{#1}}
\newcolumntype{C}[1]{>{\centering\let\newline\\\arraybackslash\hspace{0pt}}m{#1}}
\newcolumntype{R}[1]{>{\raggedleft\let\newline\\\arraybackslash\hspace{0pt}}m{#1}}
\newcolumntype{H}{>{\setbox0=\hbox\bgroup}c<{\egroup}@{}}

\SetCommentSty{mycommfont}

\renewcommand{\paragraph}[1]{\noindent\textbf{#1.}}

\ifdefmacro{\ispreprint}{%
  \makeatletter
  \newcommand\footnoteref[1]{\protected@xdef\@thefnmark{\ref{#1}}\@footnotemark}
  \makeatother
}{}

\newcommand{\addcomment}[2]{\ifdefmacro{\showcomment}{{\textcolor{#1}{#2}}}{}}
\definecolor{darkgreen}{RGB}{0,153,0}
\definecolor{dark-gray}{gray}{0.4} % Higher is lighter

\newcommand{\ben}[1]{\addcomment{blue}{[B: #1]}}

\newcommand{\michael}[1]{\addcomment{darkgreen}{[Michael: #1]}}

\ifdefmacro{\ispreprint}{}{%

}
\begin{document}

\title{On Graph Neural Network Fairness in the Presence of Heterophilous Neighborhoods}

\author{Donald Loveland}
\affiliation{%
  \institution{University of Michigan}}
\email{dlovelan@umich.edu}

\author{Jiong Zhu}
\affiliation{%
  \institution{University of Michigan}}
\email{jiongzhu@umich.edu}

\author{Mark Heimann}
\affiliation{%
  \institution{Lawrence Livermore National Lab}}
\email{mheimann@umich.edu}

\author{Ben Fish}
\affiliation{%
  \institution{University of Michigan}}
\email{benfish@umich.edu}

\author{Michael T. Schaub}
\affiliation{%
  \institution{RWTH Aachen University}}
\email{schaub@cs.rwth-aachen.de}

\author{Danai Koutra}
\affiliation{%
  \institution{University of Michigan}}
\email{dkoutra@umich.edu}

\input{PAGES/0abstract.tex}

\maketitle

\input{PAGES/1introduction.tex}
\input{PAGES/2related.tex}

\input{PAGES/3method.tex}
\input{PAGES/4results.tex}
\input{PAGES/5conclusion.tex}

\balance
\bibliographystyle{plain} 
\bibliography{main} 

\end{document}

%% file: PAGES/0abstract.tex
\begin{abstract}

We study the task of node classification for graph neural networks (GNNs) and establish a connection between group fairness, as measured by statistical parity and equal opportunity, and local assortativity, i.e., the tendency of linked nodes to have similar attributes. Such assortativity is often induced by homophily, the tendency for nodes of similar properties to connect. Homophily can be common in social networks where systemic factors have forced individuals into communities which share a sensitive attribute. Through synthetic graphs, we study the interplay between locally occurring homophily and fair predictions, finding that not all node neighborhoods are equal in this respect -- neighborhoods dominated by one category of a sensitive attribute often struggle to obtain fair treatment, especially in the case of diverging local class and sensitive attribute homophily. After determining that a relationship between local homophily and fairness exists, we investigate if the issue of unfairness can be associated to the design of the applied GNN model. We show that by adopting heterophilous GNN designs capable of handling disassortative group labels, group fairness in locally heterophilous neighborhoods can be improved by up to 25\% over homophilous designs in real and synthetic datasets. 
\end{abstract}

%% file: PAGES/1introduction.tex
\vspace{-.25cm}
\section{Introduction}

Graphs provide a natural means to represent social networks given their ability to capture relational information. Unfortunately, social networks in practice are often plagued by systemic issues leading to segregation between communities in the networks~\cite{mcpherson2001birds}. 
%Examples of sensitive attributes often segregated throughout a social network include ethnicity, gender, and social status~\cite{hofstra2017sources}. 
%\michael{again the question of what comes first: is it the sensitive attributes that lead to seggregation due to homophily? or are people becoming more alike because they are connected?}
Given the adoption of graph neural networks (GNNs) to social network based tasks, it is important to characterize the factors which reinforce harmful bias and develop strategies to promote fairness. 

In many applications, imbalance persists as a driver for discrimination due to the propensity for models to create spurious associations.
By introducing a graph structure, GNNs can amplify these associations further depending on the aggregator and local neighborhood. 
Recent work has found assortativity, the tendency of linked nodes to have similar attributes, to amplify unfairness \cite{dong2022edits, rahman2019fairwalk, spinelli2021dropout}. Often, assortativity is a byproduct of \textit{homophily}, the tendency for nodes of similar properties
to connect, which is especially true in social networks where groups have experienced prejudice.
Unfortunately, in cases where homophily has been studied, the impacts are only analyzed through a globally aggregated metric, neglecting to consider local deviations where unfairness may impact particular groups \cite{dai2021fairgnn, li2020dyadic}. That said, for predictive accuracy, particular GNN designs have been shown to improve learning in diverse settings. While early GNNs actively relied on homophily, leading to poor learning in settings where the ego-node (central node) and neighboring nodes were different, i.e. \textit{heterophilous}, recent designs have adopted more expressive aggregation procedures \cite{zhu2020homophily}. Despite their success in many applications, heterophilous GNN designs have yet to be considered to improve fair learning as the connection of homophily to fairness is understudied. 

As there are no thorough studies on how an individual's neighborhood impacts their ability to receive fair treatment, we devise a set of experiments to establish this connection. Specifically, we probe how local homophily variations in a network impacts fair treatment. As many networks experience a wide range of local homophily values, we also consider how heterophilous GNNs perform across different neighborhoods to motivate future fair GNN designs. Our experiments are performed on two real-world graphs, Pokec and NBA \cite{dai2021fairgnn}, as well as synthetic data. 
Our contributions are:
%\vspace{-.51cm}
\begin{itemize}[noitemsep,topsep=1pt] %,wide, labelwidth=!, labelindent=5pt]
    \item We design a novel synthetic graph generation strategy that enables the analysis of fairness in the context of homophily, 
    class and sensitive attributes imbalance, and feature bias (i.e. how much a feature encodes a sensitive attribute).    
    %\michael{here things become a little more precise for the first time, but some of these terms are still vague. Property imbalance? noisy encoding? what does all of this mean?}
    \item We establish the first connection between local neighborhood structure and GNN fairness, finding neighborhoods dominated by a single sensitive attribute 
    %(high homophily or high heterophily \jiong{why high sensitive attribute \emph{heterophily} means that ``neighborhood is dominated by one sensitive attribute''?}) 
    can be treated unfairly by models which assume homophily. 
    %\michael{what does fairness mean for us -- we should say it early on. Otherwise we are bound to disappoint any reader}
    \item We perform the first comparison of homophilous and heterophilous GNNs in the context of fairness, finding that heterophilous designs, on average, improve fairness at no cost to performance. 
    \item We perform the first analysis of feature bias amplification due to homophily, finding homophilous GNNs
    amplify feature bias at a greater rate
    than heterophilous GNNs.
\end{itemize}
\vspace{-.2cm}

%% file: PAGES/2related.tex
\section{Related Work}

\noindent \textbf{Evaluating Fairness.} 
Fairness is often evaluated through the lens of \textit{group} fairness or \textit{individual} fairness. Satisfying group fairness with a respect to a \emph{sensitive attribute} requires that an algorithm treat groups with different sensitive attribute values the same. In contrast, individual fairness argues that similar individuals, regardless of any particular sensitive attribute, should receive similar treatment \cite{dwork2012indivfair}. The majority of prior research in GNN fairness has focused on group fairness through measures such as statistical parity and equal opportunity. This choice can be attributed to the difficulty in designing a similarity metric which encompasses both feature and structure similarity, as well as the intricate prior knowledge required to assure the similarity metric considers systemic factors \cite{fleisher2021whatsfair}. For instance, previous work that applies individual fairness to GNNs requires a large pairwise similarity matrix defined by an expert \cite{dong2021individual}. As our focus is to investigate whether structural patterns give rise to unfairness, we focus on group fairness to characterize the difficulty in receiving preferential treatment as a community. 

\vspace{0.1cm}
\noindent \textbf{Graph Structure and Fairness.} FairWalk proposed a random walk to promote nodes from different sensitive groups at equal rates, irrespective of their group sizes \cite{rahman2019fairwalk}. Fair GNNs have adopted a similar strategy by modifying the adjacency matrix to create more balanced neighborhoods \cite{li2020dyadic, spinelli2021dropout}. In both cases, global homophily is identified as a source of unfairness and is actively changed during the optimization procedures.
That said, as GNNs often operate on a small computational neighborhood, local homophily on a k-hop neighborhood plays a more significant role. Neglecting to consider the distribution of local homophily in a network may lead to GNN design choices which reinforce bias, such as in the case of GNNs with strong homophily assumptions. Thus, it is worthwhile to consider if models which are able to reasonably learn on diverse homophily levels can also improve fairness in such neighborhoods.
%\ben{The following two sentences are unclear, need to be more clear about what our novelty is here: GNNs use local homophily, not global homophily, but previous work only shows correlation between global homophily and bias (?) so this leaves open if heterophilic-adjusted GNNs help with bias, and we should care about heterophilic GNNs rather than homophilic GNNs with a fairness constraint because...}. 

%That said, discussion on homophily is primarily used as a motivator for the proposed methods, rather than as contextualization for empirical results. Neglecting to consider natural homophily in the datasets analyzed, as well as the changes in homophily during the optimization, obscures how homophily interacts with fairness. 

\vspace{0.1cm}
\noindent \textbf{Learning in Diverse Neighborhoods.} In order to learn representations for each node in the graph, GNNs adopt an aggregation mechanism that synthesizes both the ego-node's features and the neighboring nodes' features. Depending on the neighborhood structure surrounding the ego-node, a particular aggregation mechanism may be insufficient to adequately learn representations \cite{yan2021two}. For example, GCN \cite{kipf2016semi}, GAT \cite{velickovic2018graph}, and SGC \cite{wu2019sgc} all assume homophilous neighborhoods through their weighted average of the ego- and neighboring nodes' features. To remedy this issue, methods such as GraphSAGE \cite{hamilton2017sage}, FA-GCN \cite{bo2021fagcn}, and GCN-II \cite{chen2020gcnii} decouple the ego-node from the neighbor nodes, either through a residual connection or concatenation of the ego-node embeddings. More advanced methods such as H2GCN \cite{zhu2020homophily} adopt further decoupling across higher order networks, aggregating each $k$-hop network separately and minimizing oversmoothing in highly heterophilious networks.

%% file: PAGES/3method.tex
\section{Methodology}

Our goal is to characterize the impacts of local homophily on the fairness of GNN node predictions.
In this section, we motivate and detail the methodology behind our experiments, focusing on (a) choice of evaluation, (b) the need for local neighborhood analysis, (c) synthetic dataset generation, and (d) GNN designs. The empirical evaluations are performed on two naturally occuring real-world datasets, Pokec and NBA \cite{dai2021fairgnn} outlined in Table \ref{table:datasets}. While Pokec has two subgraphs often used as benchmarks, we focus on just Pokec-n (which we refer to as Pokec) given the highly similar nature between the two. Analysis is also included on a synthetic dataset detailed in both Table \ref{table:datasets} and Section 3.3. As denoted in Table \ref{tab:datasets}, all three datasets consider the binary case for the class and sensitive attribute. The nature of these datasets is crucial as many recent works in GNN fairness have generated feature similarity graphs on tabular data which may not be representative given our knowledge of segregation based on sensitive attributes in social networks \cite{agarwal2021towards, dong2022edits}. 

% The synthetic data proposed accounts for many factors usually considered in fair GNN research, but is not all encompassing. 
% %For instance, the synthetic data does not account for features which are uncorrelated with the sensitive attribute, but predictive of the class label, as well as different amounts of homophily on a per-property basis. To introduce such capabilities to the synthetic data generation process would require multiple new parameters which may interact with the graph generation process non-trivially. 
% Thus, we also analyze datasets proposed in \cite{dai2021fairgnn} to understand how our synthetic findings translate to real-world settings. The two datasets, Pokec and NBA, represent naturally occurring social structures in two social media sites, Pokec and Twitter. \jiong{May need to elaborate more on what does the graph structure captures since it is hard to link NBA with Twitter.} The nature of these datasets is crucial as many recent works in GNN fairness have generated feature similarity graphs on tabular data which may not be representative given our knowledge of segregation based on sensitive attributes in social networks \cite{agarwal2021towards, dong2022edits}. 

\begin{table}[h]
\centering
\vspace{-0.2cm}
\caption{Real and synthetic datasets.}
\label{tab:datasets}
\vspace{-0.2cm}
\resizebox{\columnwidth}{!}{
\begin{tabular}{l@{\hskip12pt}r@{\hskip12pt}r@{\hskip12pt}rrrr}
\toprule
   \textbf{Name} & \textbf{Nodes} & \textbf{Edges} & \textbf{Classes} & \textbf{Class}  & \textbf{Sensitive}  & \textbf{Sensitive}  \\
   \textbf{} & \textbf{} & \textbf{} & \textbf{} & \textbf{Hom.} & \textbf{Attr. Hom.} & \textbf{ Attr.}  \\
\midrule
     Pokec \cite{dai2021fairgnn} & 66,569  & 729,129 & 2 & 0.75 & 0.96 &  Region \\
     NBA \cite{dai2021fairgnn} & 403 & 16,570 & 2 & 0.40 & 0.73 &  Nationality \\
     Synthetic data &  1,000 & 10,000    & 2 & Variable & Variable & N/A \\ %\tnote{1}\\
\bottomrule
\end{tabular}
}
\vspace{-0.4cm}
\label{table:datasets}
\end{table}

\subsection{Fairness Metrics}
%reminder{maybe explain a bit what these measures capture intuitively; consider adding a table with symbols + definitions}
For all the experiments proposed, predictive performance is measured by the F1-score and fairness is measured by statistical parity: $\Delta_{SP} = |P(\hat{c} = 1 | s = 1 ) - P(\hat{c} = 1 | s = 0)|$ and equal opportunity: $\Delta_{EO} = |P(\hat{c} = 1 | c = 1, s = 1 ) - P(\hat{c} = 1 | c = 1, s = 0)|$ where $\hat{c}$ is the 
% class prediction for a node, 
predicted class for a node,
$c$ is the ground truth class, and $s$ is the sensitive attribute. $\Delta_{SP}$ measures the gap in preferential treatment that may occur between two groups, but does not necessarily consider whether the individual has been deemed qualified (i.e., doesn't consider the original class). $\Delta_{EO}$ introduces additional conditioning on the class, measuring the preferential treatment gap in the context of those who have been deemed qualified by the ground-truth labels. Each provides an important perspective when considering fairness, particularly when individuals may have been systematically kept from particular treatments solely as a result of their sensitive attribute and not because of qualification.

\subsection{Global vs. Local Perspective}

Since GNN architectures operate on a small computational graph of nodes $k$ hops away from the ego-node (where $k$ is the GNN depth), global descriptors may fail to contextualize a GNN's performance if the local structure metric distribution is diverse. Thus, characterizing the distribution of local homophily across nodes in the graph enables a finer analysis of group fairness, helping to illuminate regions where unfairness may persist. As an example of how the global homophily ratio $h$ over a graph's edge set $E$ may be misleading, where 

\vspace{-0.4cm}
\begin{equation}
h = \frac{|\{(u,v) \in E | p_u = p_v\}|}{|E|}
\label{eq:homophily_ratio}
% \vspace{-0.1cm}
\end{equation}
for some property $p$, we compare it to 1-hop and 2-hop local homophily ratios for class and sensitive attribute of the Pokec dataset \cite{dai2021fairgnn}. The local homophily ratio for a node is calculated by replacing $E$ in Equation \ref{eq:homophily_ratio} with the edges from a $k$-hop neighborhood around the node. Despite the Pokec graph displaying globally homophilous tendencies in both the class and sensitive attribute %($h_s$ and $h_c$ >= 0.7) 
as seen in Table~\ref{table:datasets}, large deviations in homophily are present even in the 2-hop neighborhood. As GNNs are usually shallow (1-2 layers), inference may be dependent on an extremely heterophilous neighborhood despite the global tendency, limiting predictive capability.

\begin{figure}[h]
    \vspace{-0.3cm}
    \centering
    \includegraphics[width=0.75\columnwidth, keepaspectratio]{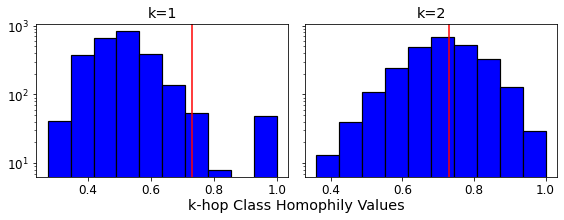} \\
    \includegraphics[width=0.75\columnwidth, keepaspectratio]{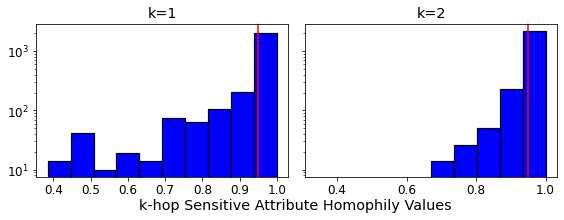}
    \vspace{-0.25cm}
    \caption{Pokec dataset: Histograms of class homophily (top) and sensitive attribute homophily (bottom) ratios in each node's k-hop neighborhood. The global homophily ratio, given in Table \ref{tab:datasets}, is denoted as a single red vertical line.}
    \vspace{-.4cm}
    \label{fig:local_vs_global_homophily}
\end{figure}

Given the ramifications that poor performance in particular regions of the graph could have on group fairness, we first investigate the variation that arises across different combinations of local class and sensitive attribute homophily ratios. Specifically, given a graph, we stratify the nodes into groups based on their surrounding (1- or 2-hop) neighborhood's class and sensitive attribute homophily ratio. Each of these groups are then analyzed in regards to their predictive performance and group fairness relative to the other groups to determine if specific homophily ratio combinations lead to unfairness. Through this knowledge, effective models and training paradigms can be established to handle poor performing regions.

\subsection{Synthetic Data Generation}

% \subsubsection{Synthetic Data}
While real-world data serves as a strong proxy for application, it can be difficult to assess how properties related to fairness, such as property skew, structural variation, and feature bias, interact.
To formalize the relationship between structure and group fairness, we propose a synthetic graph generation scheme enabling control over the previously described properties of interest. While previous preferential attachment models have only considered class homophily  ~\cite{karimi2018homophily, zhu2020homophily}, we extend the framework to allow for control over both class and sensitive attribute homophily, as well as feature bias. As is common in group fairness literature, the data is generated with a binary sensitive attribute. Final details of the generated graphs are detailed in Table \ref{table:datasets}, and generation details are outlined below.

\vspace{0.05cm}
\noindent \textbf{Attribute and Feature Generation.} Class $c$ and sensitive attribute $s$ are sampled from input joint probability distribution $P(C, S)$ where $C$ and $S$ denote the set of class labels and sensitive attribute values. Note that $P(C, S)$ can be uniform or skewed to create either class or sensitive attribute imbalance. Features $f$ are generated from a 2D Gaussian where each dimension is centered at $0 \pm (e * s)$. Parameter $e \in [0, 1]$ introduces noise into the sensitive attribute. When $e = 1$, the sensitive attribute $s$ is explicitly encoded, when $e = 0$, the sensitive attribute is removed. $e$ provides the capability to analyze scenarios where other features are correlated to the sensitive attribute, referred to as feature bias. 
%\jiong{Give some explanation on the meaning of $s$ here?}

\vspace{0.05cm}
\noindent \textbf{Compatibility Matrix.} Compatibility matrices are generated for the class and sensitive attribution, describing the probability of two nodes with certain properties connecting. We assume that the two compatibility matrices are independent. Both compatibility matrices are formally defined as $[\mathbf{H}_{\mathcal{S}}]_{u, v} = P\left((u, v) \in E | \mathrm{S}_u = s_u, \mathrm{S}_v = s_v\right) $ and $[\mathbf{H}_{\mathcal{C}}]_{u, v} = P\left((u, v) \in E | \mathrm{C}_u = c_u, \mathrm{C}_v = c_v\right) $ for two nodes $u$ and $v$. 
% User-defined $h_{c}$ and $h_{s}$ define the probability of connecting to a node of similar property (homophilous) in each matrix, and thus the diagonal entries of each matrix is set to $h_{c}$ and $h_{s}$. 
We set the diagonal entries of each matrix as $h_{c}$ and $h_{s}$ which define the probability of connecting nodes with similar class and sensitive attribution (i.e., level of homophily).
The off-diagonals in $[\mathbf{H}_{\mathcal{C}}]_{u, v}$ and $[\mathbf{H}_{\mathcal{S}}]_{u, v}$ are set to $1-h_c$ and $1-h_s$, respectively, to reflect the probability of edges connecting which are between different properties (heterophilous). 

\vspace{0.05cm}
\noindent \textbf{Structure Generation.} Given $c$, $s$, and both compatibility matrices, the probability of a newly generated node $u$ being attached to an existing node $v$ is defined as $P\left((u, v) \in E\right) \propto [\mathbf{H}_{\mathcal{S}}]_{u, v} \cdot [\mathbf{H}_{\mathcal{C}}]_{u, v} \cdot d_v$, where $d_v$ is the degree of node $v$. \\

\vspace{-0.3cm}
\subsection{Homophilous vs Heterophilous GNNs}

To facilitate fair learning across all regions of the graph, we apply the experimental set up for local homophily analysis outlined in the previous section to different GNN designs. Rather than evaluate the explicit instantiations of the homophilous and heterophilous designs outlined in Section 2, we aggregate performance across models that adopt each design to test the design itself. This provides an additional benefit of smoothing out variation from model initialization and stochastic training. In our results, we refer to a set of models designed for homophilous graphs as \textit{homophily} models. This set of models includes GCN \cite{kipf2016semi}, GAT \cite{velickovic2018graph}, and SGC \cite{wu2019sgc}. Likewise, we also include a set of models we refer to as \textit{heterophily} models which includes GCN-II \cite{chen2020gcnii}, GraphSage \cite{hamilton2017sage}, FA-GCN \cite{bo2021fagcn}, and H2GCN \cite{zhu2020homophily}. From an evaluation perspective, we compare the global performance of these two different designs, as well as differences in their performance across regions of varying homophily. 

%% file: PAGES/4results.tex
\section{Results}

Our experiments are performed over homophilous and heterophilous GNN designs. For each design type, we average across the associated models. In the NBA and Pokec experiments, we perform 10 and 3 training runs, respectively, for each model. In the synthetic experiments, we generate 10 different synthetic graphs and perform 3 training runs per model on each graph. All hyperparameters are tuned based on ranges recommended in the associated paper and the final hyperparameters are those which maximize the predictive performance on a validation test. All metrics reported are on an external test set and the train-validation-test splits (50-25-25) are kept the same during each training phase. For local evaluation, bins of size 0.2 are used to group 1-hop class homophily and 1-hop sensitive attribute homophily regions of the graph. 

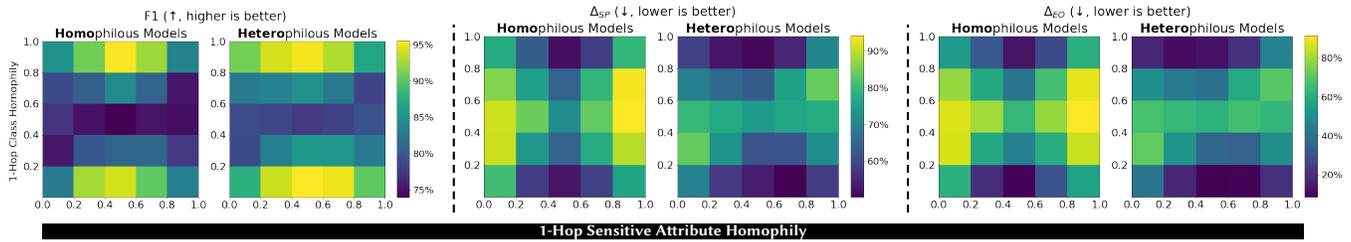
\begin{figure*}[t]
    \centering
    \includegraphics[width=0.675\columnwidth, keepaspectratio]{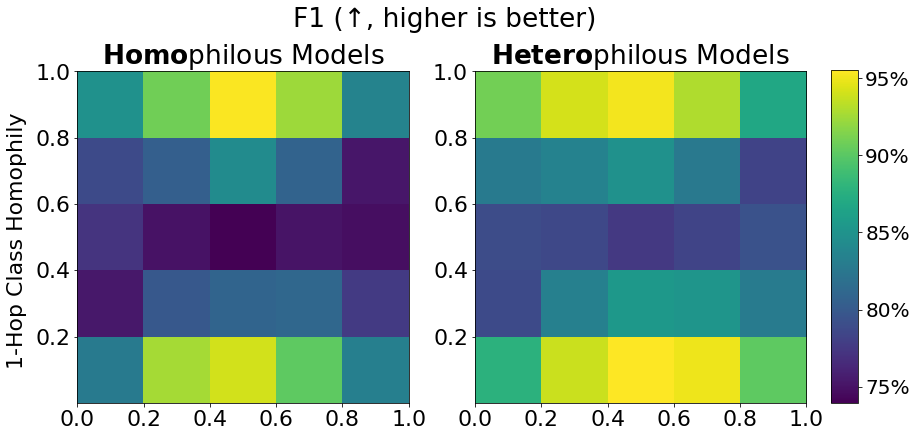} \hfill 
    \tikz{\draw[-,black, densely dashed, thick](0,-1.5) -- (0,1.0);} 
    \includegraphics[width=0.675\columnwidth, keepaspectratio]{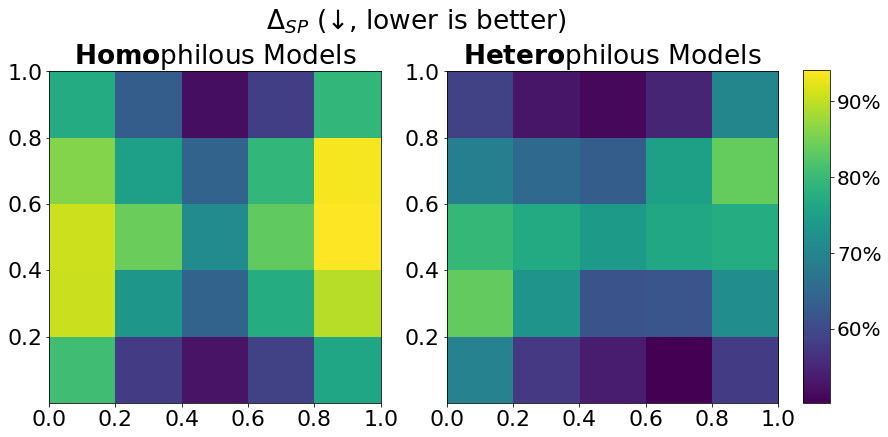} \hfill
    \tikz{\draw[-,black, densely dashed, thick](0,-1.5) -- (0,1.0);} 
    \includegraphics[width=0.675\columnwidth, keepaspectratio]{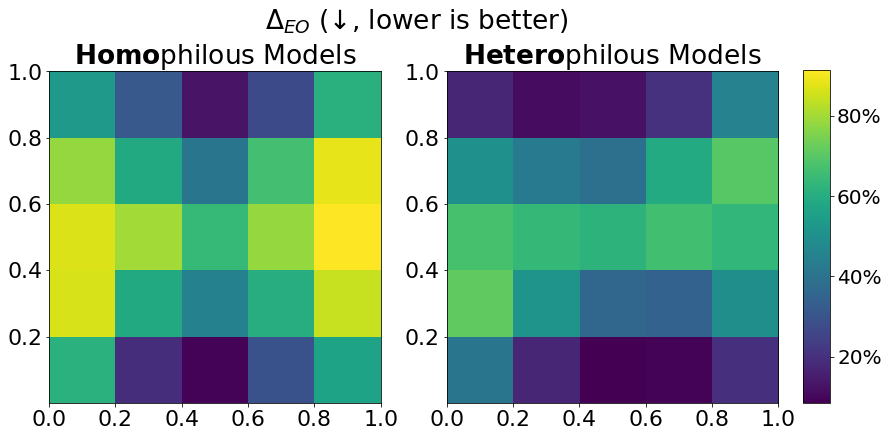}
    \begin{tikzpicture}
        \node at (0,0) [xaxis-fill] {};   
%\fill [gray] (0,0) rectangle (17,0.2) node[pos=.5] {\scriptsize 1-Hop Sensitive Attribute Homophily};
    \end{tikzpicture}
    \vspace{-.35cm}
    \caption{Synthetic data with uniform $P(C, S)$: For neighborhoods with different local sensitive attribute homophily (x axis) and local class homophily (y axis), we provide $F1$ (left), $\Delta_{SP}$ (center), and $\Delta_{EO}$ (right) results for homophilous and heterophilous models. Across each metric, heterophilous models outperform homophilous models for nearly all neighborhood configurations.}
    \vspace{-.4cm}
    \label{fig:homphily_vs_heterophily}
\end{figure*}

\subsection{Synthetic Experiments}
%\jiong{(Maybe put the takeaways as the title of each paragraph (and correspond directly to the contributions)?)}

%We begin by posing two questions: (1) How do homophily-assumed and heterophily-adjusted models perform in different neighborhoods of the synthetic graphs? (2) How much does homophily-assumed and heterophily-adjusted models amplify bias in the case that the sensitive attribute is noisy?

\begin{figure}[b]
    \vspace{-.4cm}
    \centering

    \begin{subfigure}{1.0\linewidth}
    \centering
        \includegraphics[width=0.9\columnwidth, keepaspectratio]{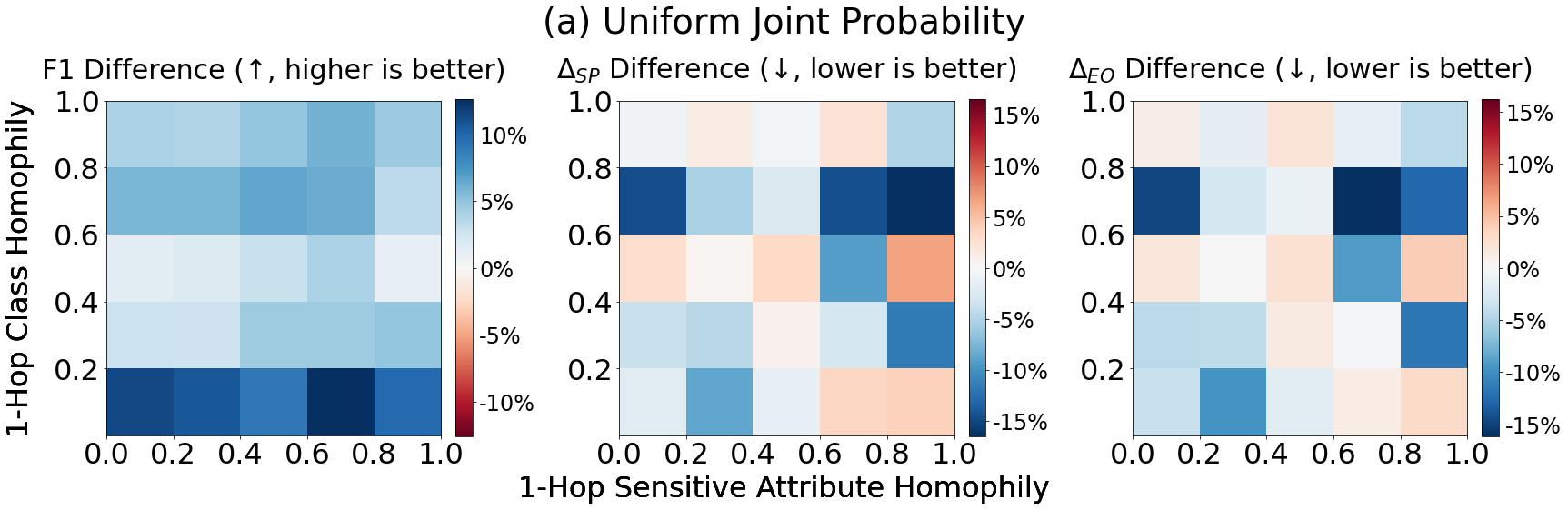} 
        %\subcaption{\small Uniform Joint Probability}
    \end{subfigure} \\
    %\includegraphics[width=1.0\columnwidth, keepaspectratio]{FIGURES/local_diff_synth-edited-notitle.png} \\
    %\subcaption{Uniform Joint Probability}
    \def\dashfill{\cleaders\hbox{--}\hfill}
    \hbox to \hsize{\dashfill\hfil}
    \begin{subfigure}{1.0\linewidth}
    \centering
        \includegraphics[width=0.9\columnwidth, keepaspectratio]{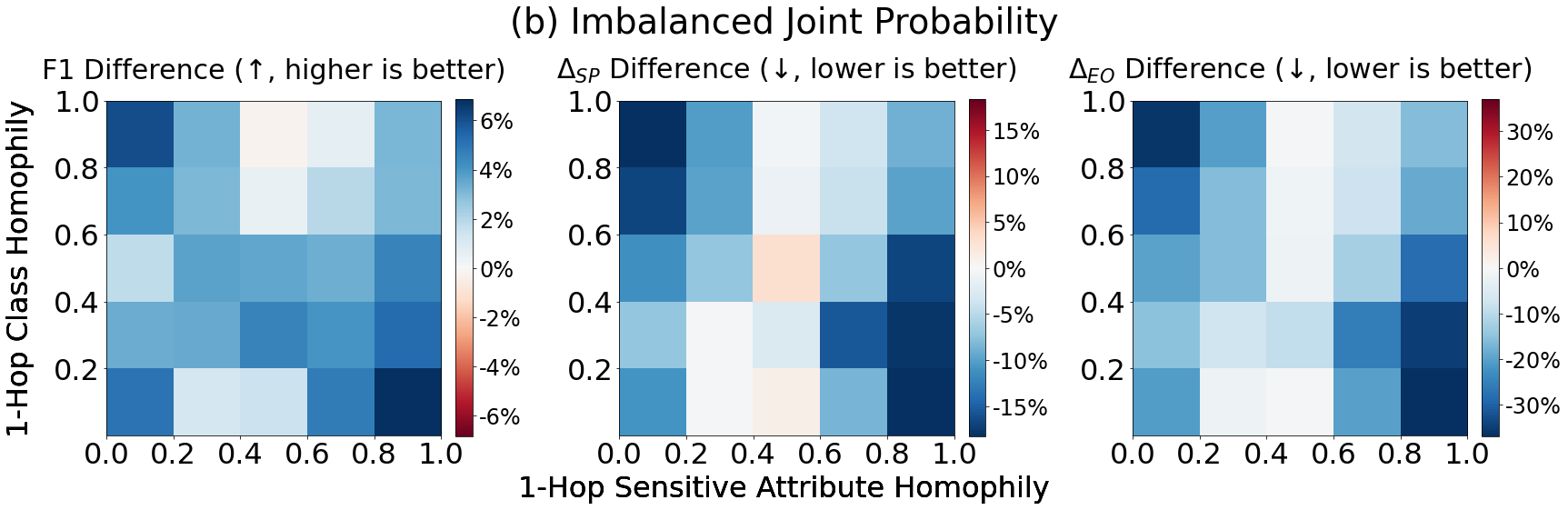}
        %\subcaption{Imbalanced Joint Probability}
    \end{subfigure}
    %\includegraphics[width=1.0\columnwidth, keepaspectratio]{FIGURES/local_diff_synth_skew-edited--notitle.png}
    %\subcaption{Imbalanced Joint Probability}
    \vspace{-0.3cm}
    \caption{
    (top) Synthetic data with Uniform $P(C, S)$: Difference between heterophilous and homophilous models for each metric. %in $F1$ (left), $\Delta_{SP}$ (center), and $\Delta_{EO}$ (right). 
    Blue indicates an improvement per metric for heterophilous models. 
    (bottom) Synthetic data with Imbalanced $P(C, S)$: We show the same results for the imbalanced case. The improvement of heterophilous over homophilous models is more significant, esp.\ for diverging local homophily.} 
    \label{fig:local_diff}
    \vspace{-.2cm}
\end{figure}

\subsubsection{\textbf{Are heterophilous GNNs more fair than homophilous GNNs?}}

Our first set of experiments generates synthetic data on each combination of class and sensitive attribute homophily values, $h_c$ and $h_{s} \in \{0.1, 0.3, 0.5, 0.7, 0.9\}$, to determine how local variations give rise to different model performance. We find that \textit{heterophilous models improve fairness over homophily-assumed models without degrading predictive performance} across nearly all neighborhood types. In Figure \ref{fig:homphily_vs_heterophily}, we plot F1, $\Delta_{SP}$, and $\Delta_{EO}$ across different neighborhood types for our sets of homophilous and heterophilous models. The synthetic graphs used exhibit no class or sensitive attribute imbalance, indicating that any group unfairness can only be a result of aggregation. We find that for nearly every neighborhood type, heterophilous models produce higher F1 score while minimizing group unfairness. This is most prevalent in regions of high sensitive attribute homophily and heterophily (i.e., $h_s \geq 0.8$ or $h_s \leq 0.2$).

This issue is further exemplified in Figure \ref{fig:local_diff} where the heterophilous model metrics are subtracted from the homophilous model metrics, 
%\mh{maybe we could get away with just visualizing differences if space is an issue?} 
demonstrating the change in performance and group fairness due to the heterophilous models. These calculations are performed for both the uniform and skewed case, where the skew in $P(C, S)$ makes an individual 3x as likely to adopt a particular class given their sensitive attribute. The imbalanced case helps further elucidate the increased group fairness in neighborhoods which are dominated by one sensitive attribute when adopting heterophilous models. Interestingly, neighborhoods which possess a discrepancy between class homophily and sensitive attribute homophily appear most impacted by homophilous designs -- $\Delta_{SP}$ and $\Delta_{EO}$ degrade by ~18\% and ~35\%, respectively. As there is a known connection between homophilous model design and neighborhood over-smoothing for predictions \cite{yan2021two}, it is possible fairness also experiences a similar phenomena when homophily over a sensitive attribute is present.

\subsubsection{\textbf{Do homophilous GNNs amplify feature bias more than heterophilous GNNs?}} 
\begin{figure}[b]
    \vspace{-.4cm}
    \centering
    \includegraphics[width=0.9\columnwidth, keepaspectratio]{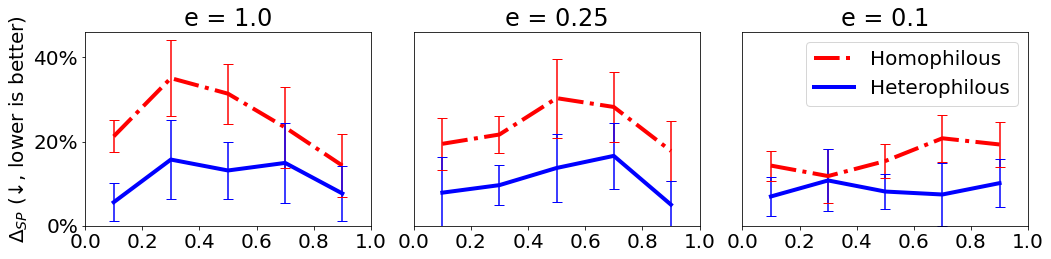}\\
    \includegraphics[width=0.9\columnwidth, keepaspectratio]{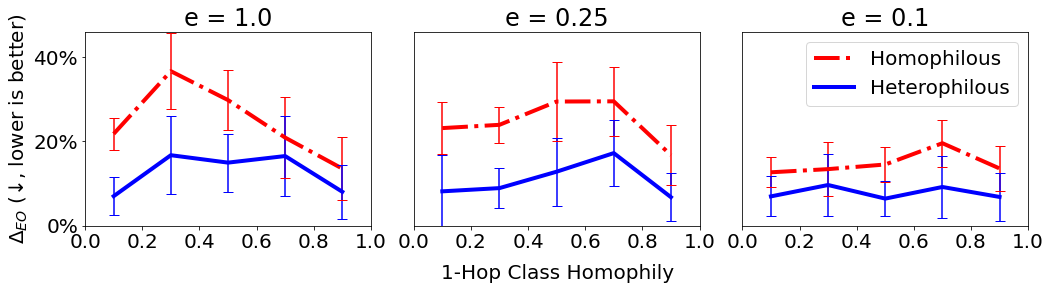}\\
    \vspace{-0.3cm}
    \caption{Synthetic data with uniform $P(C, S)$:  $\Delta_{SP}$ (top) and  $\Delta_{EO}$ (bottom) for homophilous and heterophilous models as the sensitive attribute encoding, $e$, is varied in high sensitive attribute homophily settings. %Each row shares the same y-axis range. 
    Homophilous models are shown to maintain high unfairness even with low $e$.}
    \label{fig:encoding_diff}
    %\vspace{-0.4cm}
\end{figure}

Our second set of synthetic experiments find that \textit{homophilous models amplify feature bias, even through features that are weakly correlated to the sensitive attribute}. Figure \ref{fig:encoding_diff} compares the performance of the homophilous and heterophilous models as the noise level $e$ is varied in the synthetic data. This experiment focuses on the impact of $e$ in the uniform and high sensitive attribute homophily in order to isolate the impact of aggregation. As established in Figure \ref{fig:homphily_vs_heterophily}, the left-most plot of Figure \ref{fig:encoding_diff} demonstrates that heterophily based models achieve lower unfairness in both metrics as compared to homophilous models. As $e$ becomes smaller, i.e., removing the sensitive attribute from the model, it is evident that the homophily models retain high unfairness. Even at $e$ = 0.25, we find that most of the fairness calculations for the homophily model have either stayed the same or dropped by a couple percent despite the sensitive attribute being significantly less salient. While the heterophily models also experience a similar small drop in percentage, each represents a larger relative change given heterophily models' fairness metrics are low to begin with. Given the homophily assumption present in the homophily models, we find the aggregation is able to exaggerate the sensitive attribute despite the significant noise. As demonstrated by the decrease in fairness, this amplification is able to occur despite the class label, reflecting potentially harmful situations.
%\vspace{-.4cm}

\begin{figure}[t]
    \centering
    \includegraphics[width=0.82\columnwidth, keepaspectratio]{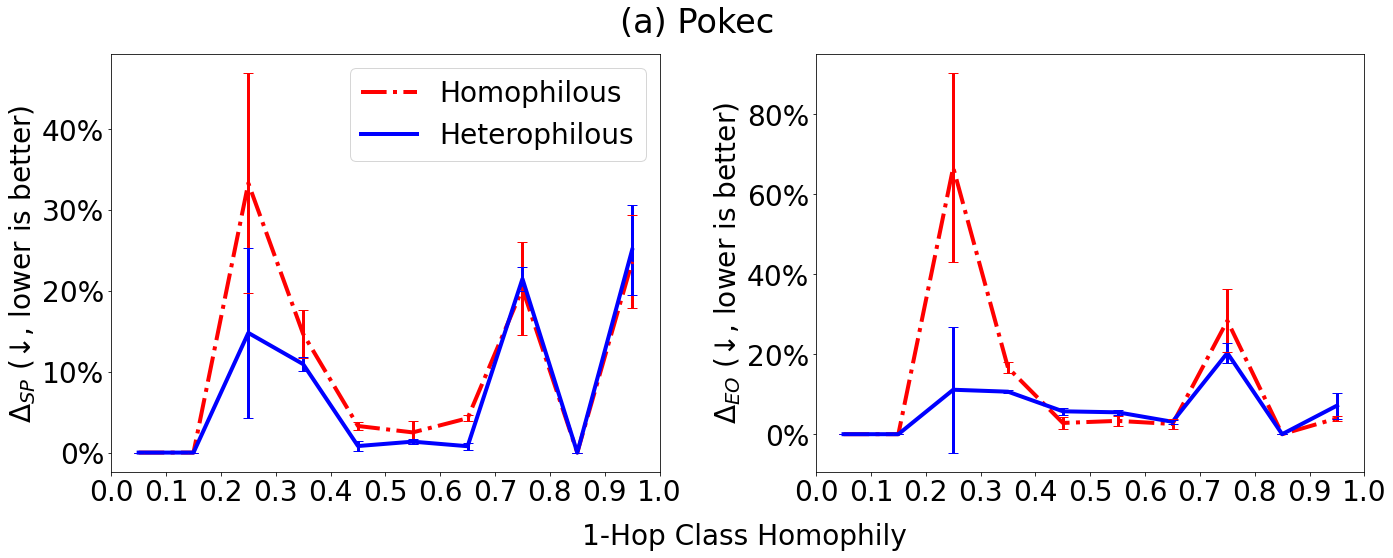}\\
    %\def\dashfill{\cleaders\hbox{--}\hfill}
    %\hbox to \hsize{\dashfill\hfil}
    \includegraphics[width=0.82\columnwidth, keepaspectratio]{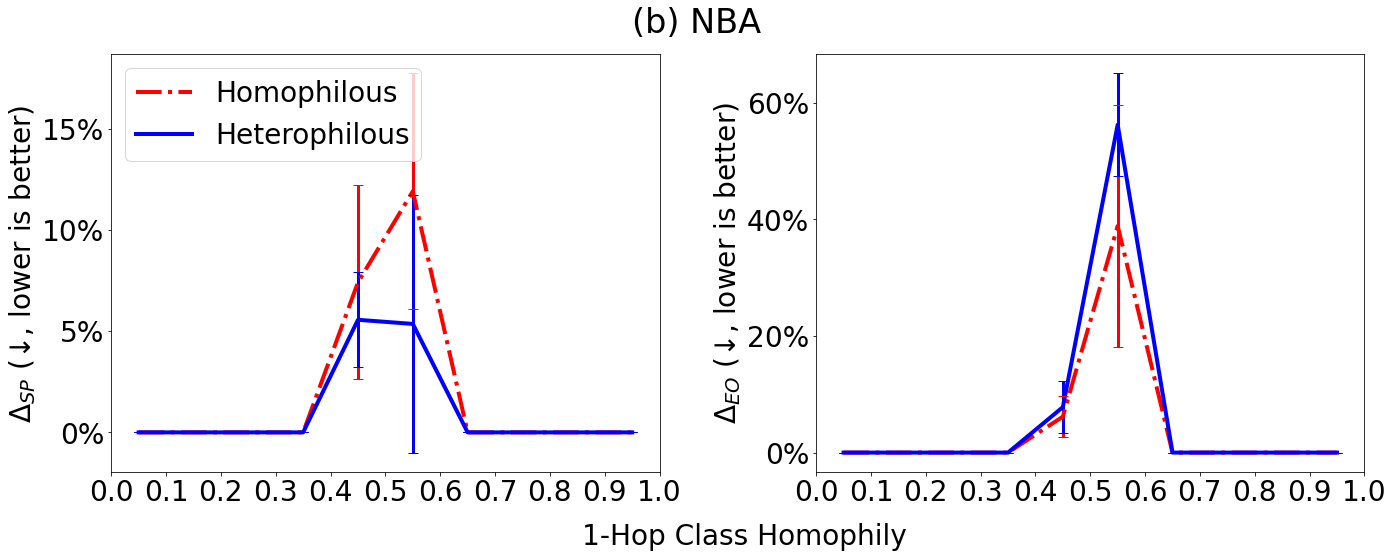}\\
    \vspace{-0.3cm}
    \caption{Variations across class homophily in $\Delta_{SP}$ and $\Delta_{EO}$ for the (top) Pokec and (bottom) NBA dataset in regions of high sensitive attribute homophily. Pokec possesses significant fairness differences in low class homophily regimes.}
    \label{fig:real-world}
    
    \vspace{-.5cm}
\end{figure}

\subsubsection{\textbf{Discussion}}
Together, our synthetic findings indicate two possible issues that may arise when analyzing the group fairness of GNN models. First, it may be the case that global group fairness metrics improve without necessarily treating everyone in the network equally. This is evident in our first experiment where individuals' placement within the network significantly impacted their ability to receive preferential treatment. %Even when accounting for those with the preferred class through $Delta_{EO}$, we find similar disparity,  
Second, it may be the case that directly removing the presence of a sensitive attribute, either through data anonymization or adversarial debiasing, will not cleanse the dataset of bias. As seen in the second experiment, if high sensitive attribute homophily persists despite low $e$, group fairness becomes poor. 

\subsection{Real-world Experiments}

\begin{table}[b]
\vspace{-0.5cm}
\caption {ACC, $\Delta_{SP}$, and $\Delta_{EO}$ on Pokec and NBA datasets}
\vspace{-0.2cm}
{\small
\begin{tabular}{@{}lllll@{}}
\toprule
       & Model Type   & ACC & $\Delta_{SP}$ & $\Delta_{EO}$ \\ \midrule
\multirow{2}{*}{Pokec}  & Homophilous   &   0.66$\pm$0.02 & 0.06$\pm$0.04   &   0.04$\pm$0.02                          \\ \cmidrule(l){2-5}
                         & Heterophilous &   0.69$\pm$0.00 & 0.01$\pm$0.01   &   0.02$\pm$0.00                            \\ \midrule                 
\multirow{2}{*}{NBA}   & Homophilous   &   0.63$\pm$0.05 & 0.08$\pm$0.05   &   0.06$\pm$0.02                            \\ \cmidrule(l){2-5}
                               & Heterophilous &   0.72$\pm$0.02 & 0.04$\pm$0.03   &   0.11$\pm$0.04                         \\ \bottomrule
\label{table:real_results}
\end{tabular}
}
\end{table}

To assess how the synthetic experiments may translate to real-world settings, we analyze Pokec and NBA. The accuracy, $\Delta_{SP}$, and $\Delta_{EO}$ are first computed on the entire graph for each model type as shown in Table \ref{table:real_results}. To understand if a model may be more likely to make unfair predictions in the regimes deemed most unfair in the synthetic data, we aggregate all high sensitive attribute homophily neighborhoods ($h_s > 0.6$), accounting for 98.5\% and 93.7\% of the nodes within the Pokec and NBA datasets, respectively and compute performance. Pokec's wide range of class homophily values illuminates the difficulty in producing fair predictions across the entire class homophily spectrum. A gap begins to emerge for neighborhoods with $h_{c} < ~0.4$, indicating different treatments when the local class and sensitive attribute homophily differ. These results agree with the findings on the synthetic datasets, particularly when probing the case of divergence between class and sensitive attribute. Unfortunately, due to NBA's extremely small graph size, it is hard to draw any conclusive results regarding fairness in both the global and local behavior. That said, heterophilous models' ability to still improve predictive performance without degrading fairness is valuable to consider during GNN model selection.

\begin{comment}

We also evaluate metrics based on equal opportunity~\cite{hardt2016equality} and demographic parity, which measure the disparity in recall rate and likelihood ratio of the ``advantaged'' outcome among all sensitive groups. 

Evaluation metric introducation: Could make sense to either introduce here if not as prevalent (we just use what others use), if more intimately tied with GNN design, introduce with methodology. 
\begin{itemize}
        \item dataset introduction -- include size used in experiments, homophily global, something about local distribution (point back to local vs global)
        \item Localized analysis of heterophily impact on fairness and prediction for synthetic and real data
        \item FIG: how homo vs hetero models perform across stratified regions of homophily in class and sensitive attribute
        \item Noisy encoding of the sensitive attribute
        \item NOT YET DONE: stratifying results on class, are those with preferential treatment vs non-preferential treatment treated differently -- may be achieved with EO 
\end{itemize}

\begin{itemize}
    \item Reiterate final results
    \item ground in possible situations of unfairness -- high sens attr homophily but low class homophily could represent individuals who are experiencing some form of social segregation but were able to break out and achieve preferential treatment. due this likely being a minority situation it is unclear what will be reinforced snd this work offers the first analysis on what may occur  
\end{itemize}
\end{comment}

%% file: PAGES/5conclusion.tex
\section{Conclusion}

In this work, we study how local structure around a node in the form of homophily can influence a GNN's fairness. In a set of synthetic experiments, we find that a) GNNs designed for homophily can produce significant unfairness in regions dominated by one sensitive attribute, b) GNNs designed for heterophily can minimize these over-generalizations and reduce unfairness in the context of $\Delta_{SP}$ and $\Delta_{EO}$, and c) data anonymization and debiasing solutions may be insufficient as noisy proxies for the sensitive attribute (feature bias) can be amplified by homophilous architectures. We test these findings further on natural real-world datasets, determining similar trends hold, particularly in the extreme homophily settings. 

\section*{Acknowledgments}
This work is supported by the National Science Foundation under CAREER Grant No.~IIS 1845491, Army Young Investigator Award No.~W911NF1810397, and Adobe, Amazon, and Google faculty awards.
Any opinions or conclusions 
% or recommendations 
expressed are those of the authors and do not reflect the views of the funding parties.